\pdfoutput=1
\documentclass[12pt,a4paper]{article}

\usepackage{ifthen} 
\newboolean{pdflatex}
\setboolean{pdflatex}{true} 

\newboolean{articletitles}
\setboolean{articletitles}{true} 

\newboolean{uprightparticles}
\setboolean{uprightparticles}{false} 

\usepackage{microtype}
\usepackage{lineno}  
\usepackage{xspace} 

\usepackage{graphicx}  
\usepackage{color}
\usepackage{colortbl}
\graphicspath{{./figs/}} 

\usepackage{amsmath} 
\usepackage{amssymb}
\usepackage{amsfonts}
\usepackage{upgreek} 
\usepackage{units}

\newcommand*\patchAmsMathEnvironmentForLineno[1]{
\expandafter\let\csname old#1\expandafter\endcsname\csname #1\endcsname
\expandafter\let\csname oldend#1\expandafter\endcsname\csname
end#1\endcsname
 \renewenvironment{#1}
   {\linenomath\csname old#1\endcsname}
   {\csname oldend#1\endcsname\endlinenomath}
}
\newcommand*\patchBothAmsMathEnvironmentsForLineno[1]{
  \patchAmsMathEnvironmentForLineno{#1}
  \patchAmsMathEnvironmentForLineno{#1*}
}
\AtBeginDocument{
\patchBothAmsMathEnvironmentsForLineno{equation}
\patchBothAmsMathEnvironmentsForLineno{align}
\patchBothAmsMathEnvironmentsForLineno{flalign}
\patchBothAmsMathEnvironmentsForLineno{alignat}
\patchBothAmsMathEnvironmentsForLineno{gather}
\patchBothAmsMathEnvironmentsForLineno{multline}
}

\usepackage{hyperref}    
\usepackage[all]{hypcap} 

\def\lhcb {\mbox{LHCb}\xspace}
\def\ux85 {\mbox{UX85}\xspace}

\ifthenelse{\boolean{uprightparticles}}
{

 \def\Pmu         {\ensuremath{\upmu}\xspace}                 
 \def\Pnu         {\ensuremath{\upnu}\xspace}                 
                  
 \def\Ppi         {\ensuremath{\uppi}\xspace}

 \def\Ppsi        {\ensuremath{\uppsi}\xspace}

 \def\PDelta      {\ensuremath{\Delta}\xspace}                 
 \def\PXi      {\ensuremath{\Xi}\xspace}                 
 \def\PLambda      {\ensuremath{\Lambda}\xspace}                 
 \def\PSigma      {\ensuremath{\Sigma}\xspace}                 
 \def\POmega      {\ensuremath{\Omega}\xspace}                 
 \def\PUpsilon      {\ensuremath{\Upsilon}\xspace}

 \def\PB      {\ensuremath{\mathrm{B}}\xspace}                 
                  
 \def\PD      {\ensuremath{\mathrm{D}}\xspace}

 \def\PJ      {\ensuremath{\mathrm{J}}\xspace}                 
 \def\PK      {\ensuremath{\mathrm{K}}\xspace}

 \def\Pb      {\ensuremath{\mathrm{b}}\xspace}                 
 \def\Pc      {\ensuremath{\mathrm{c}}\xspace}                 
 \def\Pd      {\ensuremath{\mathrm{d}}\xspace}

 \def\Pi      {\ensuremath{\mathrm{i}}\xspace}

 \def\Ps      {\ensuremath{\mathrm{s}}\xspace}

}
{

 \def\Pmu         {\ensuremath{\mu}\xspace}                 
 \def\Pnu         {\ensuremath{\nu}\xspace}                 
                  
 \def\Ppi         {\ensuremath{\pi}\xspace}

 \def\Ppsi        {\ensuremath{\psi}\xspace}                 
                  
 \mathchardef\PDelta="7101
 \mathchardef\PXi="7104
 \mathchardef\PLambda="7103
 \mathchardef\PSigma="7106
 \mathchardef\POmega="710A
 \mathchardef\PUpsilon="7107
                  
 \def\PB      {\ensuremath{B}\xspace}                 
                  
 \def\PD      {\ensuremath{D}\xspace}

 \def\PJ      {\ensuremath{J}\xspace}                 
 \def\PK      {\ensuremath{K}\xspace}

 \def\Pb      {\ensuremath{b}\xspace}                 
 \def\Pc      {\ensuremath{c}\xspace}                 
 \def\Pd      {\ensuremath{d}\xspace}

 \def\Pi      {\ensuremath{i}\xspace}

 \def\Ps      {\ensuremath{s}\xspace}

}

\def\mup        {\ensuremath{\Pmu^+}\xspace}
\def\mun        {\ensuremath{\Pmu^-}\xspace}

\def\neu        {\ensuremath{\Pnu}\xspace}

\def\neum       {\ensuremath{\neu_\mu}\xspace}

\def\dquark    {\ensuremath{\Pd}\xspace}

\def\squark    {\ensuremath{\Ps}\xspace}

\def\cquark    {\ensuremath{\Pc}\xspace}

\def\bquark    {\ensuremath{\Pb}\xspace}

\def\pion  {\ensuremath{\Ppi}\xspace}

\def\pip   {\ensuremath{\pion^+}\xspace}
\def\pim   {\ensuremath{\pion^-}\xspace}

\def\kaon  {\ensuremath{\PK}\xspace}

  \def\Kbar  {\kern 0.2em\overline{\kern -0.2em \PK}{}\xspace}

\def\Kz    {\ensuremath{\kaon^0}\xspace}
\def\Kzb   {\ensuremath{\Kbar^0}\xspace}
\def\KzKzb {\ensuremath{\Kz \kern -0.16em \Kzb}\xspace}
\def\Kp    {\ensuremath{\kaon^+}\xspace}
\def\Km    {\ensuremath{\kaon^-}\xspace}

\def\KpKm  {\ensuremath{\Kp \kern -0.16em \Km}\xspace}

\def\Kstarz  {\ensuremath{\kaon^{*0}}\xspace}

  \def\Dbar    {\kern 0.2em\overline{\kern -0.2em \PD}{}\xspace}
\def\D       {\ensuremath{\PD}\xspace}

\def\Dz      {\ensuremath{\D^0}\xspace}
\def\Dzb     {\ensuremath{\Dbar^0}\xspace}
\def\DzDzb   {\ensuremath{\Dz {\kern -0.16em \Dzb}}\xspace}
\def\Dp      {\ensuremath{\D^+}\xspace}
\def\Dm      {\ensuremath{\D^-}\xspace}

\def\DpDm    {\ensuremath{\Dp {\kern -0.16em \Dm}}\xspace}

\def\Dstarm  {\ensuremath{\D^{*-}}\xspace}

\def\Dsm     {\ensuremath{\D^-_\squark}\xspace}

\def\B       {\ensuremath{\PB}\xspace}

  \def\Bbar    {\kern 0.18em\overline{\kern -0.18em \PB}{}\xspace}

\def\Bz      {\ensuremath{\B^0}\xspace}
\def\Bzb     {\ensuremath{\Bbar^0}\xspace}
\def\Bu      {\ensuremath{\B^+}\xspace}

\def\Bp      {\ensuremath{\Bu}\xspace}

\def\Bd      {\ensuremath{\B^0}\xspace}
\def\Bs      {\ensuremath{\B^0_\squark}\xspace}
\def\Bsb     {\ensuremath{\Bbar^0_\squark}\xspace}
\def\Bdb     {\ensuremath{\Bbar^0}\xspace}

\def\jpsi     {\ensuremath{{\PJ\mskip -3mu/\mskip -2mu\Ppsi\mskip 2mu}}\xspace}

  \def\Y#1S{\ensuremath{\PUpsilon{(#1S)}}\xspace}

\def\Lbar {\ensuremath{\kern 0.1em\overline{\kern -0.1em\PLambda}}\xspace}

\newcommand{\decay}[2]{\ensuremath{#1\!\to #2}\xspace}         

\def\to                 {\ensuremath{\rightarrow}\xspace}

\newcommand{\dms}{\ensuremath{\Delta m_{\squark}}\xspace}
\newcommand{\dmd}{\ensuremath{\Delta m_{\dquark}}\xspace}

\def\BsToJPsiPhi  {\decay{\Bs}{\jpsi\phi}}
\def\BdToJPsiKst  {\decay{\Bd}{\jpsi\Kstarz}}

\def\AT#1     {\ensuremath{A_{\mathrm{T}}^{#1}}\xspace}

\def\C#1      {\ensuremath{\mathcal{C}_{#1}}\xspace}                       
\def\Cp#1     {\ensuremath{\mathcal{C}_{#1}^{'}}\xspace}                    
\def\Ceff#1   {\ensuremath{\mathcal{C}_{#1}^{\mathrm{(eff)}}}\xspace}        
\def\Cpeff#1  {\ensuremath{\mathcal{C}_{#1}^{'\mathrm{(eff)}}}\xspace}       
\def\Ope#1    {\ensuremath{\mathcal{O}_{#1}}\xspace}                       
\def\Opep#1   {\ensuremath{\mathcal{O}_{#1}^{'}}\xspace}

\newcommand{\tev}{\ensuremath{\mathrm{\,Te\kern -0.1em V}}\xspace}
\newcommand{\gev}{\ensuremath{\mathrm{\,Ge\kern -0.1em V}}\xspace}
\newcommand{\mev}{\ensuremath{\mathrm{\,Me\kern -0.1em V}}\xspace}
\newcommand{\kev}{\ensuremath{\mathrm{\,ke\kern -0.1em V}}\xspace}
\newcommand{\ev}{\ensuremath{\mathrm{\,e\kern -0.1em V}}\xspace}
\newcommand{\gevc}{\ensuremath{{\mathrm{\,Ge\kern -0.1em V\!/}c}}\xspace}
\newcommand{\mevc}{\ensuremath{{\mathrm{\,Me\kern -0.1em V\!/}c}}\xspace}
\newcommand{\gevcc}{\ensuremath{{\mathrm{\,Ge\kern -0.1em V\!/}c^2}}\xspace}
\newcommand{\gevgevcccc}{\ensuremath{{\mathrm{\,Ge\kern -0.1em V^2\!/}c^4}}\xspace}
\newcommand{\mevcc}{\ensuremath{{\mathrm{\,Me\kern -0.1em V\!/}c^2}}\xspace}

\def\mum  {\ensuremath{\,\upmu\rm m}\xspace}

\def\invfb   {\ensuremath{\mbox{\,fb}^{-1}}\xspace}

\def\invps{\ensuremath{{\rm \,ps^{-1}}}\xspace}

\def\gsim{{~\raise.15em\hbox{$>$}\kern-.85em
          \lower.35em\hbox{$\sim$}~}\xspace}
\def\lsim{{~\raise.15em\hbox{$<$}\kern-.85em
          \lower.35em\hbox{$\sim$}~}\xspace}

\def\sWeights{\mbox{\em sWeights}}

\def\pt         {\mbox{$p_{\rm T}$}\xspace}

\def\evtgen     {\mbox{\textsc{EvtGen}}\xspace}
\def\pythia     {\mbox{\textsc{Pythia}}\xspace}

\def\geant      {\mbox{\textsc{Geant4}}\xspace}

\def\photos     {\mbox{\textsc{Photos}}\xspace}

\def\tell1  {TELL1\xspace}
\def\ukl1   {UKL1\xspace}

\usepackage{cite} 
\usepackage{mciteplus}

\newcommand{\BdToDmpip}{\decay{\Bd}{\Dm \pip}}

\begin{document}

\renewcommand{\thefootnote}{\fnsymbol{footnote}}
\setcounter{footnote}{1}

\begin{titlepage}
\pagenumbering{roman}

\vspace*{-1.5cm}
\centerline{\large EUROPEAN ORGANIZATION FOR NUCLEAR RESEARCH (CERN)}
\vspace*{1.5cm}
\hspace*{-0.5cm}
\begin{tabular*}{\linewidth}{lc@{\extracolsep{\fill}}r}
\ifthenelse{\boolean{pdflatex}}
{\vspace*{-2.7cm}\mbox{\!\!\!\includegraphics[width=.14\textwidth]{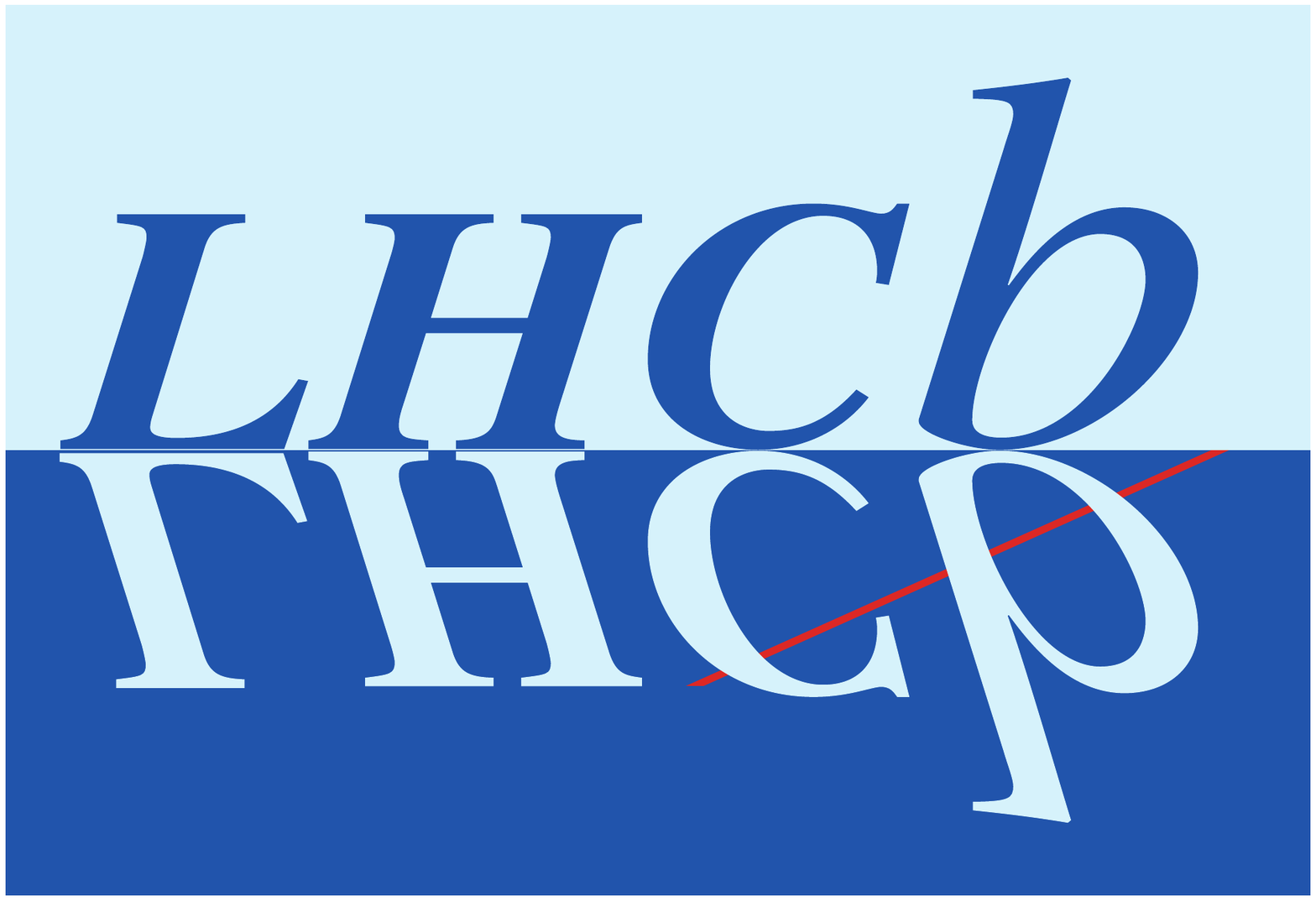}} & &}
{\vspace*{-1.2cm}\mbox{\!\!\!\includegraphics[width=.12\textwidth]{lhcb-logo.eps}} & &}
\\
 & & CERN-PH-EP-2012-315 \\  
 & & LHCb-PAPER-2012-032 \\  
& & March 23, 2013 \\ 
 & & \\

\end{tabular*}

\vspace*{3cm}

{\bf\boldmath\huge
\begin{center}
Measurement of the \Bd--\Bdb oscillation frequency \dmd with the decays \BdToDmpip and \BdToJPsiKst
\end{center}
}

\vspace*{1.2cm}

\begin{center}
The LHCb collaboration\footnote{Authors are listed on the following pages.}
\end{center}

\vspace{1.2cm}

\begin{abstract}
  \noindent

The $\Bz$--$\Bzb$ oscillation frequency \dmd is measured by the LHCb experiment using a dataset corresponding to an integrated luminosity of $\unit[1.0]{fb^{-1}}$ of proton-proton collisions at $\sqrt{s} = \unit[7]{TeV}$, and is found to be $\dmd = 0.5156 \pm 0.0051\,(\mathrm{stat.}) \pm 0.0033\,(\mathrm{syst.})\invps$. The measurement is based on results from analyses of the decays \BdToDmpip ($\Dm\to\Kp\pim\pim$) and \BdToJPsiKst ($\jpsi\to\mup\mun$, $\Kstarz\to\Kp\pim$) and their charge conjugated modes.
\end{abstract}

\vspace*{1.2cm}

\begin{center}
Published in Physics Letters B 719 (2013), pp. 318-325
\end{center}

\vspace{\fill}

\end{titlepage}

\newpage
\setcounter{page}{2}
\mbox{~}
\newpage

\centerline{\large\bf LHCb collaboration}
\begin{flushleft}
\small
R.~Aaij$^{38}$, 
C.~Abellan~Beteta$^{33,n}$, 
A.~Adametz$^{11}$, 
B.~Adeva$^{34}$, 
M.~Adinolfi$^{43}$, 
C.~Adrover$^{6}$, 
A.~Affolder$^{49}$, 
Z.~Ajaltouni$^{5}$, 
J.~Albrecht$^{35}$, 
F.~Alessio$^{35}$
% ...

{\footnotesize \it
% ...
$ ^{5}$Clermont Universit\'{e}, Universit\'{e} Blaise Pascal, CNRS/IN2P3, LPC, Clermont-Ferrand, France\\
$ ^{6}$CPPM, Aix-Marseille Universit\'{e}, CNRS/IN2P3, Marseille, France\\
% ...
$ ^{11}$Physikalisches Institut, Ruprecht-Karls-Universit\"{a}t Heidelberg, Heidelberg, Germany\\
% ...
$ ^{33}$Universitat de Barcelona, Barcelona, Spain\\
$ ^{34}$Universidad de Santiago de Compostela, Santiago de Compostela, Spain\\
$ ^{35}$European Organization for Nuclear Research (CERN), Geneva, Switzerland\\
% ...
$ ^{38}$Nikhef National Institute for Subatomic Physics, Amsterdam, The Netherlands\\
% ...
$ ^{43}$H.H. Wills Physics Laboratory, University of Bristol, Bristol, United Kingdom\\
% ...
$ ^{49}$Oliver Lodge Laboratory, University of Liverpool, Liverpool, United Kingdom\\
% ...
$ ^{n}$LIFAELS, La Salle, Universitat Ramon Llull, Barcelona, Spain\\
% ...
}
\end{flushleft}

\cleardoublepage

\renewcommand{\thefootnote}{\arabic{footnote}}
\setcounter{footnote}{0}

\pagestyle{plain} 
\setcounter{page}{1}
\pagenumbering{arabic}

\section{Introduction}
\label{sec:Introduction}
The frequency \dmd of oscillations between \Bz mesons and \Bzb mesons also describes the mass difference \dmd between the physical eigenstates in the \Bz--\Bzb system, and has been measured at LEP \cite{external:dmd_lep}, the Tevatron \cite{external:dmd_D0, external:dmd_CDF}, and the $B$ factories  \cite{external:dmd_Babar, external:dmd_Belle}. The current world average is $\dmd = \unit[0.507\pm 0.004]{ps^{-1}}$ \cite{external:PDG}, whilst the best single measurement prior to this Letter is by the Belle experiment, $\dmd = \unit[0.511\pm 0.005\,(\mathrm{stat.}) \pm 0.006\,(\mathrm{syst.})]{ps^{-1}}$ \cite{external:dmd_Belle}. In this document the convention $\hbar = c = 1$ is used for all units.

With increasing accuracy of the measurement of \dms, the counterpart of \dmd in the \Bs --\Bsb system \cite{internal:dms}, a more precise knowledge of \dmd becomes important, as the ratio  $\Delta m_d/\Delta m_s$ together with input from lattice QCD calculations \cite{external:dmddms, external:dmddmstheory} constrains the apex of the CKM unitarity triangle \cite{external:CKMFit, Bona:2006ah}. Therefore, the measurement of  \dmd provides an important test of the Standard Model\cite{external:bsm_susy, external:bsm_xhiggs}. Furthermore, \dmd is an input parameter in the determination of $\sin2\beta$ at LHCb \cite{LHCb:sin2betaconf}.

This Letter presents a measurement of \dmd, using a dataset corresponding to $\unit[1.0]{fb^{-1}}$ of $pp$ collisions at $\sqrt{s} = \unit[7]{TeV}$, using the decay channels \BdToDmpip ($\Dm\to\Kp\pim\pim$) and \BdToJPsiKst ($\jpsi\to\mup\mun$, $\Kstarz\to\Kp\pim$) and their charge conjugated modes.

For a measurement of \dmd, the flavour of the $\Bd$ meson at production and decay must be known. The flavour at decay is determined in both decay channels from the charge of the final state kaon; contributions from suppressed $\Bd\to\Dp\pim$ amplitudes are negligible. The determination of the flavour at production is achieved by the flavour tagging algorithms which are described in more detail in Sect.~\ref{sec:tagging}.

The \Bd meson is defined as unmixed (mixed) if the production flavour is equal (not equal) to the  flavour at decay. With this knowledge, the oscillation frequency \dmd of the \Bd meson can be determined using the time dependent mixing asymmetry
\begin{align}
	\label{equ:asymmetry}
	\mathcal{A}_\mathrm{mix}^\mathrm{signal}(t) = \frac{N_{\text{unmixed}}(t)-N_{\text{mixed}}(t)}{N_{\text{unmixed}}(t)+N_{\text{mixed}}(t)}
	= \cos (\dmd t),
\end{align}
where $t$ is the $\Bz$ decay time and $N_\text{(un)mixed}$ is the number of (un)mixed events.
\section{Experimental setup and datasets}
\label{sec:Detector}
The \lhcb detector~\cite{external:detector} is a single-arm forward
spectrometer covering the \mbox{pseudorapidity} range $2<\eta <5$, designed
for the study of particles containing \bquark or \cquark quarks. The
detector includes a high precision tracking system consisting of a
silicon-strip vertex detector surrounding the $pp$ interaction region,
a large-area silicon-strip detector located upstream of a dipole
magnet with a bending power of about $4$\,Tm, and three stations
of silicon-strip detectors and straw drift-tubes placed
downstream. The combined tracking system has a momentum resolution
$\Delta p/p$ that varies from $0.4$\,\% at $5$\gev to $0.6$\,\% at 100\gev,
and an impact parameter (IP) resolution of $20$\mum for tracks with high
transverse momentum. Charged hadrons are identified using two
ring-imaging Cherenkov detectors. Photon, electron and hadron
candidates are identified by a calorimeter system consisting of
scintillating-pad and pre-shower detectors, an electromagnetic
calorimeter and a hadronic calorimeter. Muons are identified by a
system composed of alternating layers of iron and multiwire
proportional chambers. The trigger consists of a hardware stage, based
on information from the calorimeter and muon systems, followed by a
software stage which applies a full event reconstruction.

Events including \BdToDmpip decays are required to have tracks with high 
transverse momentum \pt to pass the hardware trigger.
The software trigger requires a two-, three- or four-track
  secondary vertex with a large sum of the \pt of
  the tracks, significant displacement from the associated primary vertex (PV), and
   at least one track with $\pt > 1.7\gev$ and a large impact parameter
  with respect to that PV, and
  a good track fit. A multivariate algorithm is used
  for the identification of the secondary
  vertices~\cite{LHCb:trigger}.
  
Events in the decay \BdToJPsiKst are first required to pass a hardware trigger
  which selects a single muon with $\pt>1.48\gev$.  In
  the subsequent software trigger~\cite{LHCb:trigger}, at least
  one of the final state particles is required to have
  $\pt>0.8\gev$ and a large IP with respect to all
  PVs in the
  event. Finally, the tracks of two or more of the final state
  particles are required to form a vertex which is significantly
  displaced from the PVs in the event.

For the simulation studies, $pp$ collisions are generated using
\pythia~6.4~\cite{Sjostrand:2006za} with a specific \lhcb
configuration~\cite{LHCb-PROC-2010-056}.  Decays of hadronic particles
are described by \evtgen~\cite{Lange:2001uf} in which final state
radiation is generated using \photos~\cite{Golonka:2005pn}. The
interaction of the generated particles with the detector and its
response are implemented using the \geant
toolkit~\cite{Allison:2006ve, Agostinelli:2002hh} as described in
Ref.~\cite{LHCb-PROC-2011-006}.
\section{Selection}
\label{sec:selection}

The decay time $t$ of a \Bz candidate is evaluated from the measured momenta and from a vertex fit that constrains the \Bz candidate to originate from the associated PV \cite{Hulsbergen2005566}, and using $t = \ell \cdot m(\Bz) / p$, with the flight distance $\ell$. The associated PV is the primary vertex that is closest to the decaying \Bz meson. No mass constraints on the intermediate resonances are applied. For the calculation of the invariant mass $m$, no mass constraints are used in the \BdToDmpip channel, while the \jpsi mass is constrained to the world average \cite{external:PDG} in the analysis of the decay \BdToJPsiKst.

All kaons, pions and muons are required to have large \pt and well reconstructed tracks and vertices. In addition to this, particle identification is used to distinguish between pion, kaon and proton tracks.

The \BdToDmpip selection requires that the \Dm reconstructed mass be in a range of $\pm \unit[100]{MeV}$ around the world average \cite{external:PDG}. Furthermore, the \Dm decay vertex is required to be downstream of the PV associated to the \Bz candidate.

The sum of the \Dm and \pip \pt must be larger than $\unit[5]{GeV}$. The \Bz candidate invariant mass must be in the interval $\unit[5000]{} \leq m(\Kp\pim\pim\pip) < \unit[5700]{MeV}$. Additionally, the cosine of the pointing angle between the \Bz momentum vector and the line segment between PV and secondary vertex is required to be larger than $0.999$.

Candidates are classified by a boosted decision tree (BDT)~\cite{Breiman,
  Roe} with the AdaBoost algorithm\cite{AdaBoost}. The BDT is trained with $\Bs\to\Dsm\pip$ candidates with no particle ID criteria applied to the daughter pions and kaons. The cut on the BDT classifier is optimised in order to maximise the significance of the \BdToDmpip signal. Several input variables are used: the IP significance, the flight distance perpendicular to the beam axis, the vertex quality of the \Bz and the \Dm candidate, the angle between the \Bz momentum and the line segment between PV and \Bz decay vertex, the angle between the \Dm momentum and the line segment between PV and the \Dm decay vertex, the angle between the \Dm momentum and the line segment between the \Bz decay vertex and \Dm decay vertex, the IP and \pt of the \pip track, and the angle between the \pip momentum and the line segment between PV and \Bz decay vertex. Only \Bz candidates with a decay time $t > \unit[0.3]{ps}$ are accepted.

To suppress potential background from misidentified kaons in $\Dsm\to\Km\Kp\pim$ decays, all \Dm candidates are removed if they have a daughter pion candidate that might pass a loose kaon selection and are within a $\pm\unit[25]{MeV}$ mass window (the \Dm mass resolution is smaller than \unit[10]{MeV}) around the \Dsm mass when that pion is reconstructed under the kaon mass hypothesis.

Remaining background comes from $\Bz\to\Dm\rho^+$ and $\Bz\to\Dstarm\pip$ decays. In both cases the final state is similar to the signal, except for an additional neutral pion that is not reconstructed. This leads to two additional peaking components with invariant masses lower than those of the signal candidates. Therefore, for the measurement of \dmd only candidates with an invariant mass in the range $\unit[5200 \leq m < 5450]{MeV}$ are used.

The \BdToJPsiKst selection requires that the \Kstarz candidate has a $\pt>\unit[2]{GeV}$ and $\unit[826]{} \leq m(\Kp\pim) < \unit[966]{MeV}$. 

\begin{figure}[!htb]
\centering
\includegraphics[width=0.49\textwidth]{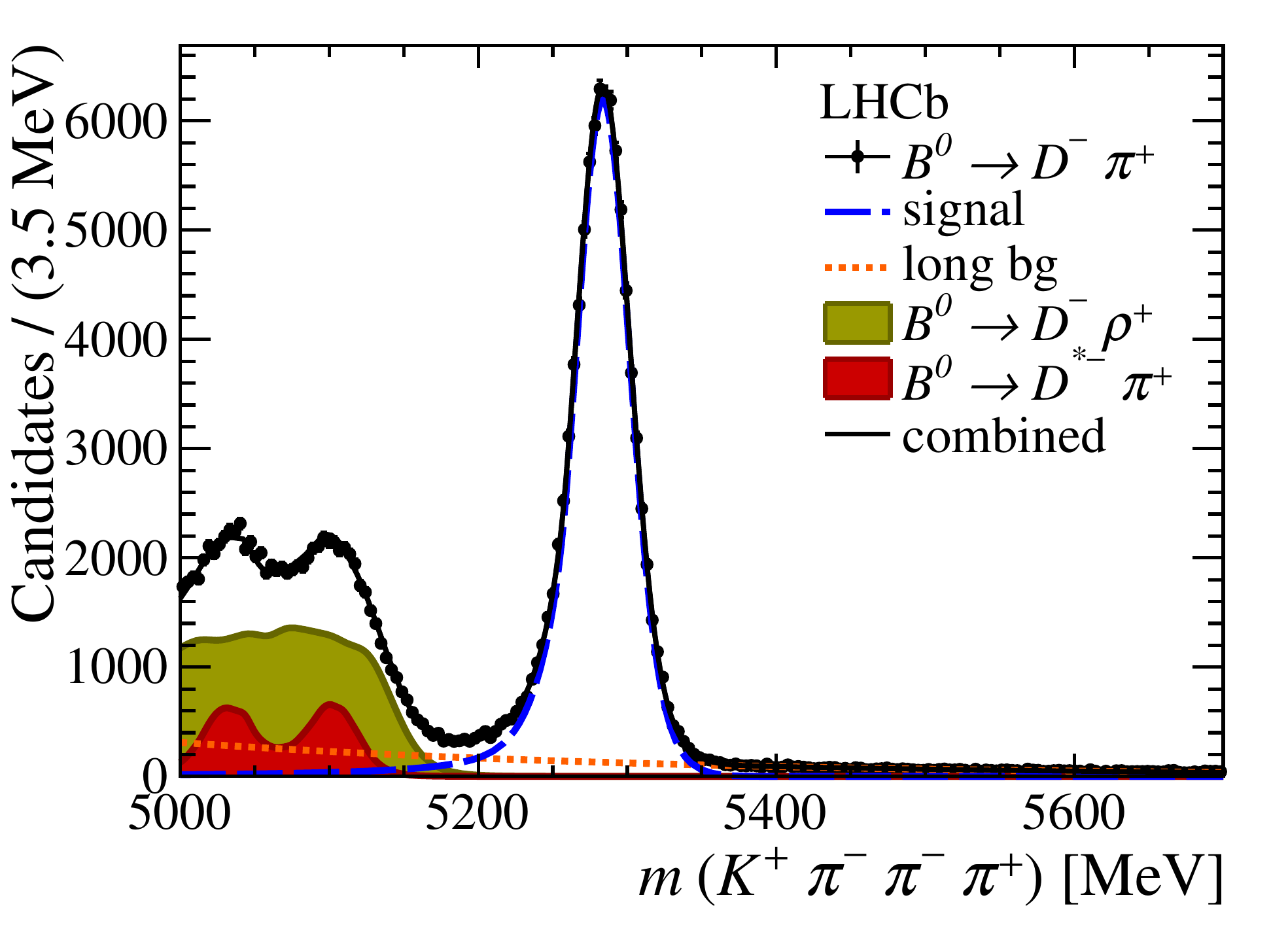}
\includegraphics[width=0.49\textwidth]{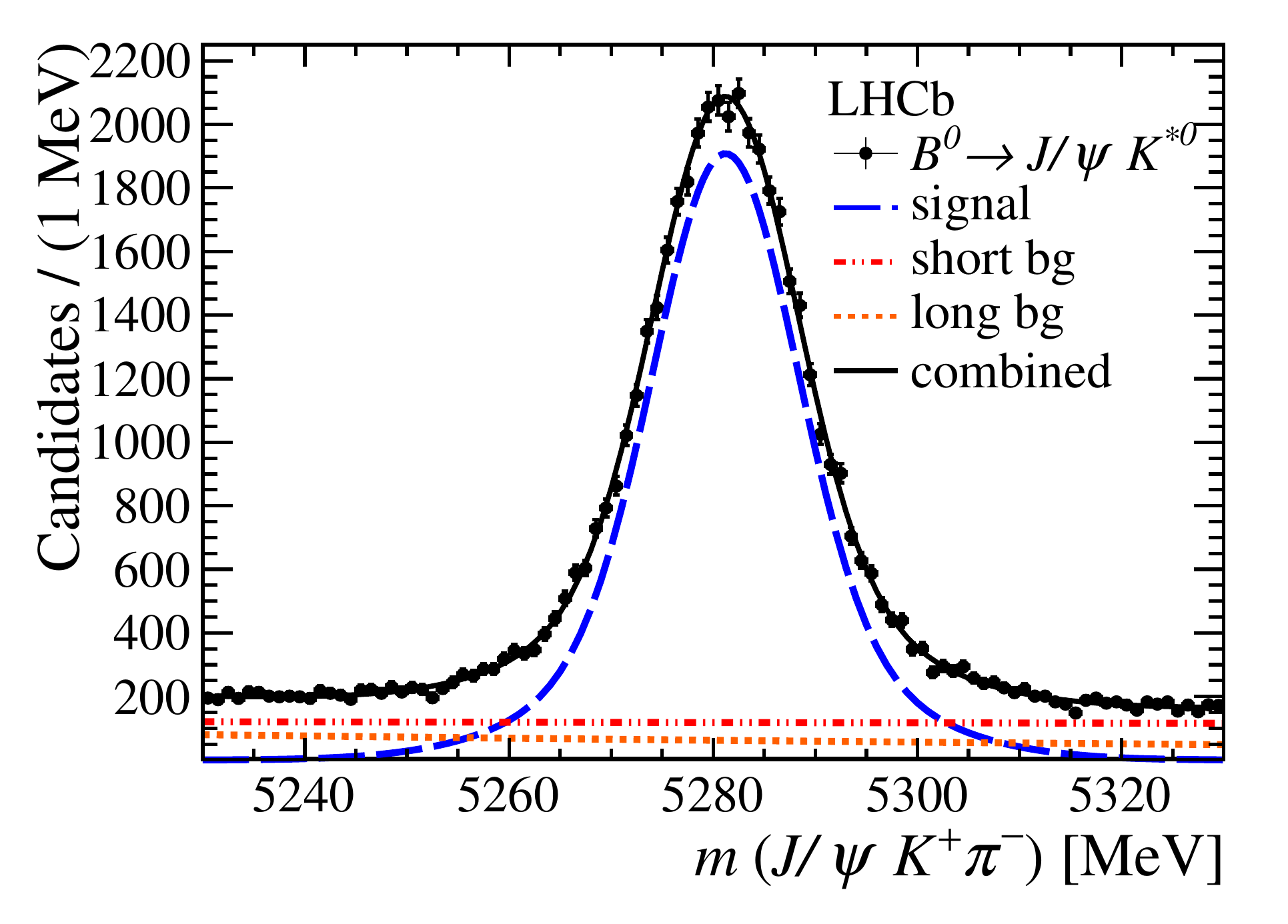}
\caption{Distribution of the \Bz candidate mass (black points). (Left) \BdToDmpip candidates with the invariant mass pdf as described in Sect.~\ref{sec:fitmodel} and two additional components for the physics background taken from MC simulated events.The blue dashed line shows the fit projection of the signal, the dotted orange line corresponds to the combinatorial background, the filled areas represent the physics background, and the black solid line corresponds to the fit projection. (Right) \BdToJPsiKst candidates, with the results of the fits described in Sect.~\ref{sec:fitmodel} superimposed. The blue dashed line shows the fit projection of the signal, the dotted orange line corresponds to the combinatorial background with long lifetime and the dash dotted red line shows the combinatorial background with short lifetime. The black solid line corresponds to the fit projection.}
\label{fig:Mass}
\end{figure}

The unconstrained $\mup\mun$ invariant mass must be within $\pm\unit[80]{MeV}$ of the \jpsi mass \cite{external:PDG}. \Bz candidates are required to have a large IP with respect to other PVs in the event and the \Bz decay vertex must be significantly separated from the PV. Additionally, \Bz candidates are required to have a reconstructed decay time $t > \unit[0.3]{ps}$ and an invariant mass in the range $\unit[5230]{} \leq m(\jpsi \Kp\pim) < \unit[5330]{MeV}$. To suppress potential background from misidentified \BsToJPsiPhi decays, all candidates are removed for which the $\Kp\pim$ mass is within a $\pm\unit[10]{MeV}$ window around the nominal $\phi(1020)$ mass when computed under the kaon mass hypothesis for the pion. The resulting mass distributions for the two decay channels are shown in Fig.~\ref{fig:Mass}.

\section{Flavour tagging}
\label{sec:tagging}

This analysis makes use of a combination of opposite side taggers and the same side pion tagger to determine the flavour of the \Bz meson at production. The opposite side taggers, which use decay products of the $b$ quark not belonging to the signal decay, are described in detail in Ref.\cite{internal:tagging}.

The same side pion tagger uses the charge of a pion that originates from the fragmentation process of the $\Bd$ meson or from decays of charged excited $B$ mesons. Pion tagging candidates are required to fulfil criteria on \pt and particle identification, as well as their IP significance and the difference between the \Bz candidate mass and the combined mass of the \Bz candidate and the pion \cite{Grabalosa:1456804}.

Depending on the tagging decision, a mixing state $q$ is assigned to each candidate, to distinguish the unmixed ($q=+1$) from the mixed ($q=-1$). Untagged events ($q=0$) are not used in this analysis. The tag and its predicted wrong tag probability $\eta_c$ are evaluated for each event using a neural network calibrated and optimized on $\Bp\to\jpsi\Kp$, \BdToJPsiKst and $\Bd\to\Dstarm\mup\neum$ events.

To take into account a possible difference in the overall tagging performance between the calibration channels and the decay channels used in this analysis, the corrected wrong tag probability $\omega$ assigned to each event is parametrised as a linear function of $\eta_c$ (the method is described and tested in Ref.~\cite{internal:tagging})
\begin{align}
  \label{equ:etac}
	\omega(\eta_{c}|p_0,p_1) = p_0 + p_1 (\eta_{c}- \langle\eta_c\rangle), 
\end{align}
where $p_0$ and $p_1$ are free parameters in the fit for \dmd described in Sect.~\ref{sec:fitmodel}. In this way, uncertainties due to the overall calibration of the tagging performance are absorbed in the statistical uncertainty on \dmd returned by the fit.
\section{Decay time resolution and acceptance}
\label{sec:acceptance}

The  decay time resolution of the detector is around $\unit[0.05]{ps}$ \cite{LHCb:phis}. This is small compared to the \Bd oscillation period of about $\unit[12]{ps}$ and does not have significant impact  on the measurement of \dmd. The resolution is accounted for by convolving a Gaussian function $G(t;\sigma_t)$, using a fixed width $\sigma_t = \unit[0.05]{ps}$, with the signal probability density function (PDF) from Eq.~(\ref{eq:timePDF}). Possible systematic uncertainties introduced by the resolution are discussed in Sect.~\ref{sec:systematics}. 

Trigger, reconstruction and selection criteria introduce efficiency effects that depend on the decay time. 
While these effects cancel in the asymmetry of Eq.~(\ref{equ:asymmetry}) for signal events, they can be important for event samples that include background. As will be shown in Sect.~\ref{sec:fitmodel}, the only relevant background in the \Bz signal region is combinatorial in nature. For this background the asymmetry $N_{q=1}^{\mathrm{bkg}}(t) - N_{q=-1}^{\mathrm{bkg}}(t)$ is expected to cancel to first order as $q$ has no physical meaning. Therefore,
\begin{align}
\label{equ:asymmetry_full}
	\mathcal{A}_\mathrm{mix}(t) &\propto \frac{(N_{q=1}^{\mathrm{sig}}(t)+N_{q=1}^{\mathrm{bkg}}(t))-(N_{q=-1}^{\mathrm{sig}}(t)+N_{q=-1}^{\mathrm{bkg}}(t))}{(N_{q=1}^{\mathrm{sig}}(t)+N_{q=1}^{\mathrm{bkg}}(t))+(N_{q=-1}^{\mathrm{sig}}(t)+N_{q=-1}^{\mathrm{bkg}}(t))}\\ \nonumber&\propto  \frac{S(t)}{S(t)+B(t)} \cos (\dmd t),
\end{align}
where $N_{q=\pm1}^{\mathrm{sig,bkg}}(t)$ denotes the number of unmixed or mixed signal (sig) and background (bkg) events. $S(t)$ and $B(t)$ denote the number of signal and background events as a function of the decay time.  Thus, the shapes of $S(t)$ and $B(t)$ have to be known to account for the time dependent amplitude of the asymmetry function.

In the analysis of decays \BdToJPsiKst, the decay time acceptance is determined from data, using a control sample of \BdToJPsiKst events that is collected without applying any of the decay time biasing selection criteria. The decay time acceptance is evaluated in bins of $t$ and is implemented in the fit described in Sect.~\ref{sec:fitmodel}.

In the decay \BdToDmpip there is no control dataset that can be used to measure the decay time acceptance. From an analysis of simulated events, it is determined that the decay time acceptance can be described by the empirical function
\begin{align}
  \label{eq:timeacc}
	\epsilon_\mathrm{acc}(t|a_1,a_2) = \arctan(a_1 \exp(a_2 t)),
\end{align}
where the parameters $a_1$ and $a_2$ are both free in the maximum likelihood fit for \dmd described in Sect.~\ref{sec:fitmodel}.

\section{Measurement of \boldmath\dmd}
\label{sec:fitmodel}
The value of \dmd is measured using a multi-dimensional extended maximum likelihood fit. The \BdToDmpip data are described by a two component PDF in which one component describes the signal and the other describes the combinatorial background. The signal component consists of the sum of a Gaussian function and a Crystal Ball function \cite{external:crystalball} with a common mean for the mass distribution, multiplied by a function $\mathcal{P}_\mathrm{sig}^t$ to describe the decay time distribution,
\begin{align}
  \label{eq:timePDF}
  &\mathcal{P}_\mathrm{sig}^t(t,q;\tau,\dmd,\omega,\sigma_t,a_1,a_2) \propto\nonumber\\ &\left[\Theta(t-\unit[0.3]{ps}) \cdot \mathrm{e}^{-\frac{t}{\tau}}\left( 1+ q (1-2\omega(\eta_{c}|p_0,p_1)) \cos\left(\Delta m_d t\right)\right)\otimes G(t;\sigma_t) \right]\nonumber\\&\cdot\epsilon_\mathrm{acc}(t|a_1,a_2).
\end{align}
Here, $\Theta(t)$ is the step function, while the \Bz lifetime $\tau$ is a free fit parameter and the average decay time resolution $\sigma_t$ is fixed. Other fit parameters are $a_1$ and $a_2$ from the decay time acceptance function $\epsilon_\mathrm{acc} (t|a_1,a_2)$ described in Sect.~\ref{sec:acceptance}, as well as the parameters $p_0$ and $p_1$ from the tagging calibration function $\omega(\eta_{c}|p_0,p_1)$ described in Sect.~\ref{sec:tagging}. Any \Bz/\Bzb production asymmetry cancels in the mixing asymmetry function, and is neglected in this analysis.

The combinatorial background component consists of an exponential PDF describing the mass distribution and the decay time PDF 
\begin{align}
	 \label{eq:timebgPDF}
  &\mathcal{P}_\mathrm{bkg}^t(t,q;\tau_\text{bkg},\omega_\text{bkg},\sigma_t) \propto\nonumber\\
 & \left[\Theta(t-\unit[0.3]{ps}) \cdot \mathrm{e}^{-\frac{t}{\tau_\text{bkg}}}\left( 1+ q (1-2\omega_\text{bkg})\right)\otimes G(t;\sigma_t) \right] .
\end{align}
The PDF is similar to the signal decay time PDF with \dmd fixed to zero. The parameter $\omega_\text{bkg}$ allows the PDF to reflect a possible asymmetry in the number of events tagged with $q=\pm 1$ in the background.  The effective lifetime $\tau_\text{bkg}$ of the long-lived background component is allowed to vary independently in the fit.

Possible backgrounds from misidentified or partially reconstructed decays are studied using mass templates determined from simulation. These are found to be negligible in the mass window $\unit[5200]{} \leq m(\Kp\pim\pim\pip) < \unit[5450]{MeV}$ that is used in the fit (c.f. Fig.~\ref{fig:Mass}).

In the \BdToJPsiKst analysis, the signal mass distribution is modelled by a double Gaussian function with a common mean and the decay time PDF is the same as described in Eq.~(\ref{eq:timePDF}), except for the decay time acceptance $\epsilon_\mathrm{acc}(t|a_1,a_2)$ that is replaced by the acceptance histogram described in Sect.~\ref{sec:acceptance} and has no free parameters. The mass distribution of the combinatorial background in \BdToJPsiKst decays is also described by an exponential function. However, the decay time distribution includes a second component of shorter lifetime to account for prompt \jpsi candidates passing the selection. The long-lived component is described by the same function as the combinatorial background in \BdToDmpip decays as in Eq.~(\ref{eq:timebgPDF}), whereas the short-lived component is described by a simple
exponential function. No other significant source of background is found.

\begin{figure}[!htb]
\centering
\includegraphics[width=0.49\textwidth]{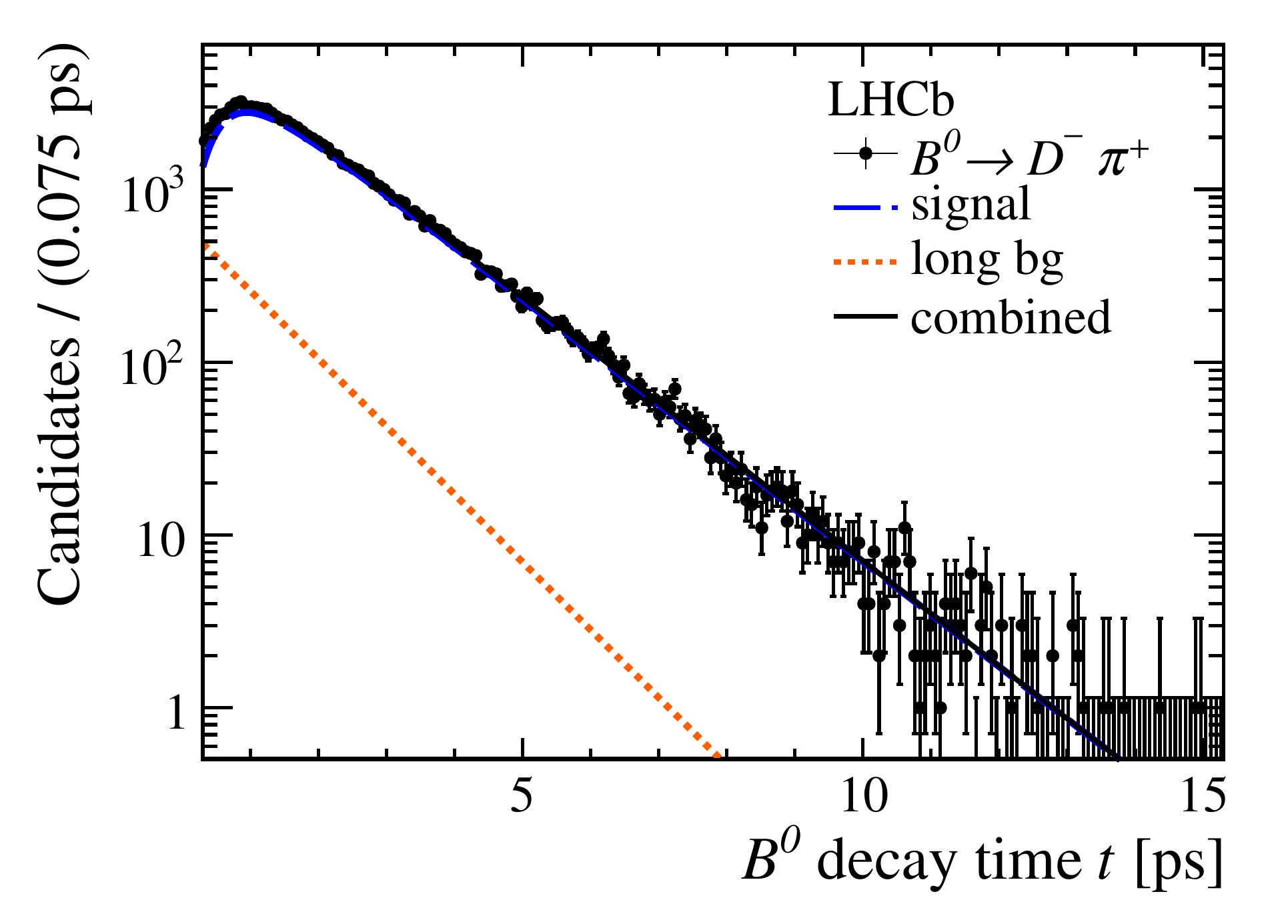}
\includegraphics[width=0.49\textwidth]{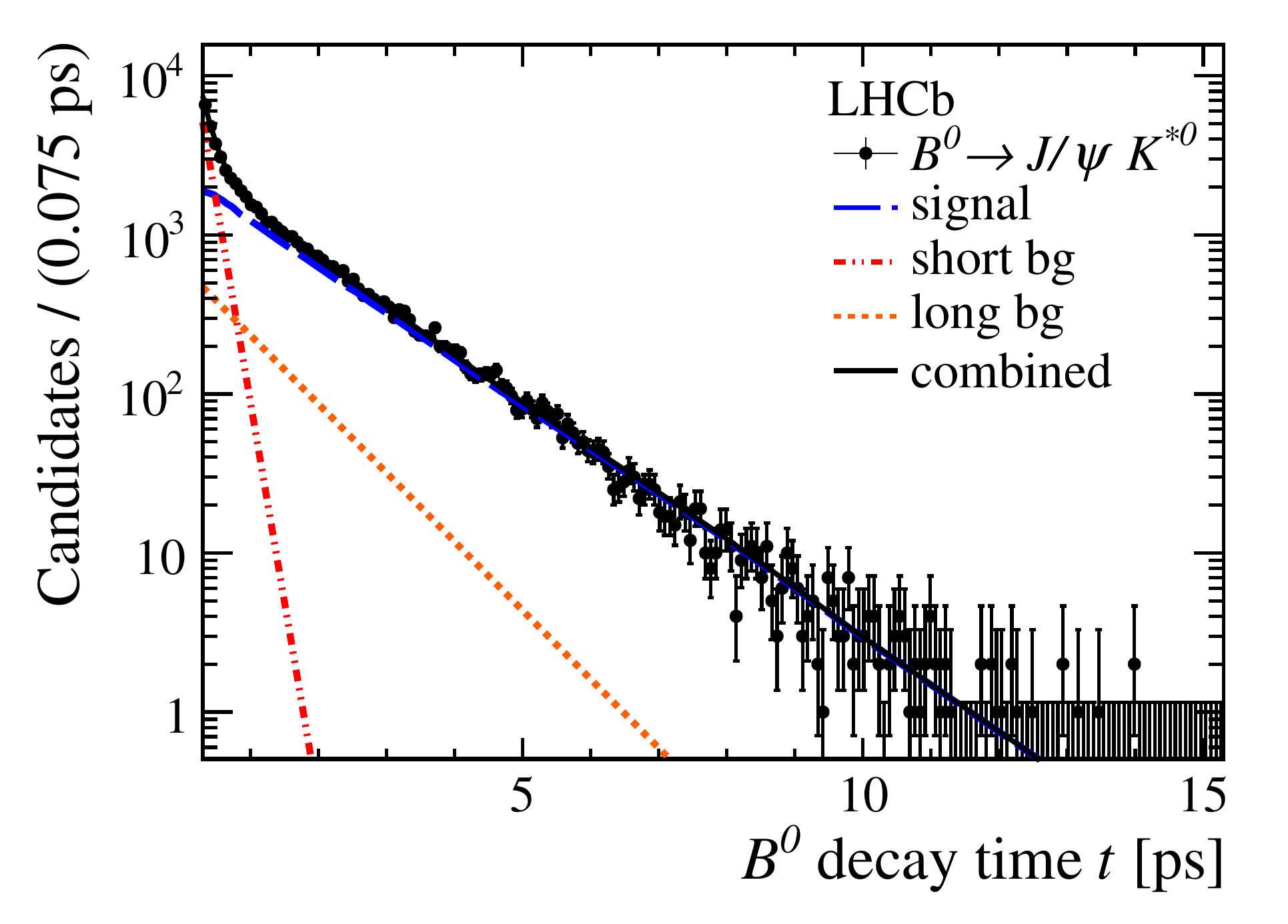}
\caption{Distribution of the decay time (black points) for (left) \BdToDmpip and (right) \BdToJPsiKst candidates. The blue dashed line shows the fit projection of the signal, the dotted orange line corresponds to the combinatorial background with long lifetime and the dash dotted red line shows the combinatorial background with short lifetime (only in the \BdToJPsiKst mode). The black solid line corresponds to the projection of the combined PDF.}
\label{fig:Time}
\end{figure}

\begin{figure}[!htb]
\centering
\includegraphics[width=0.49\textwidth]{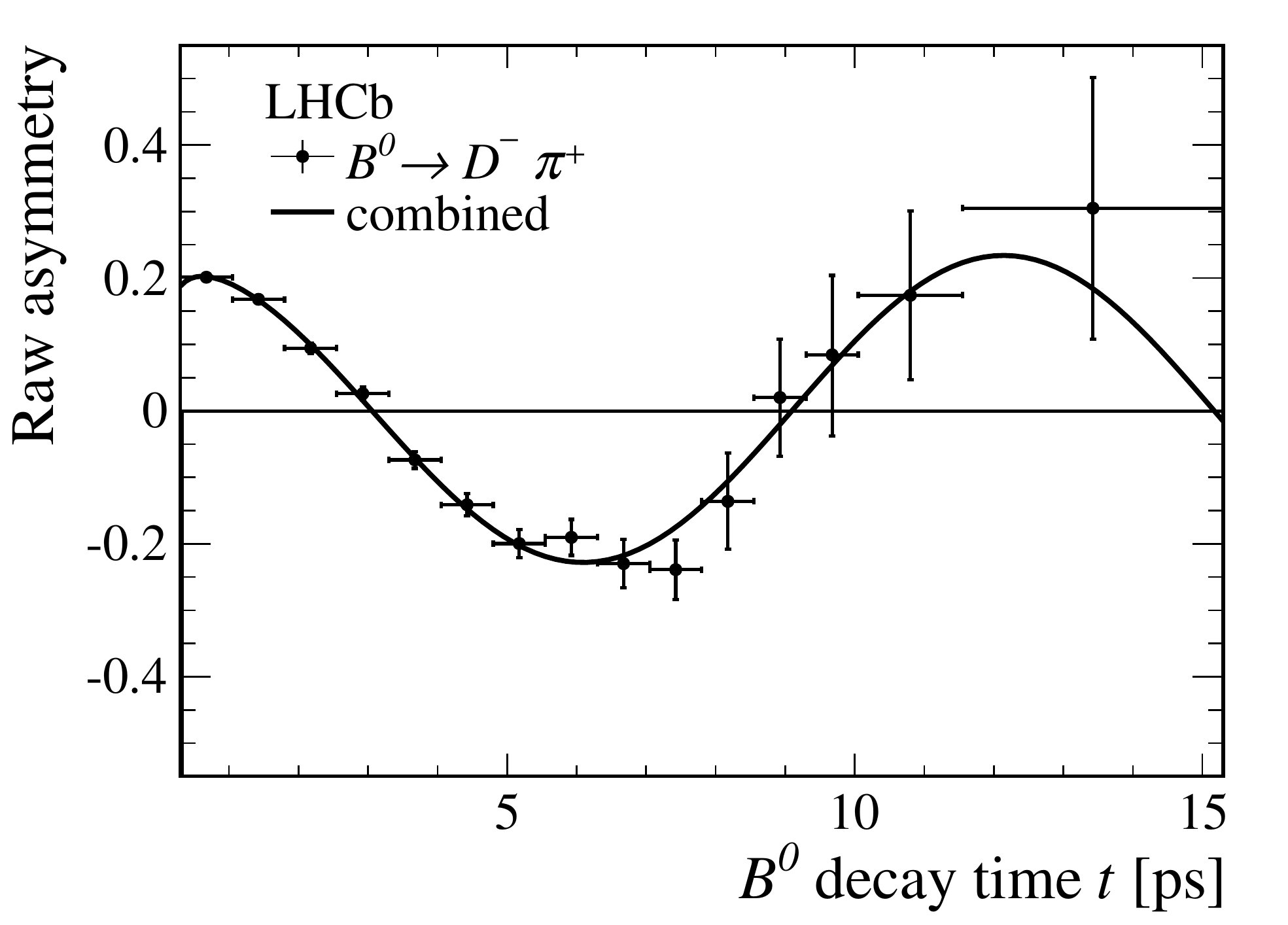}
\includegraphics[width=0.49\textwidth]{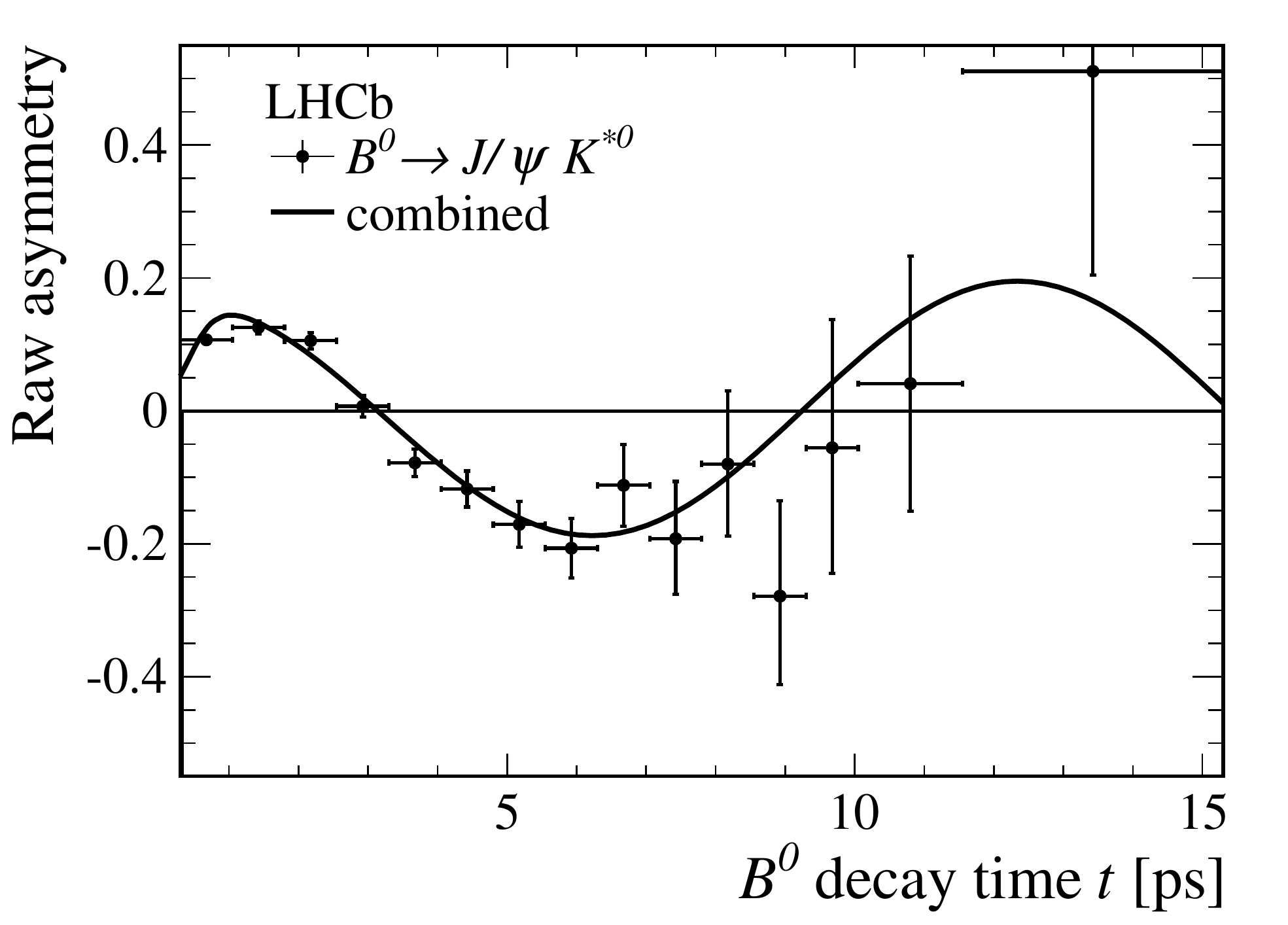}
\caption{Raw mixing asymmetry $\mathcal{A}_\mathrm{mix}$ (black points) for (left) \BdToDmpip  and (right) \BdToJPsiKst candidates. The solid black line is the projection of the mixing asymmetry of the combined PDF.}
\label{fig:Asym}
\end{figure}

The resulting values for \dmd are $\unit[0.5178 \pm 0.0061]{ps^{-1}}$ and $\unit[0.5096 \pm 0.0114]{ps^{-1}}$ in the \BdToDmpip and \BdToJPsiKst decay modes respectively. The fit yields $87\,724 \pm 321$ signal decays for \BdToDmpip and $39\,148 \pm 316$ signal decays for \BdToJPsiKst. The fit projections onto the decay time distributions are displayed in Fig.~\ref{fig:Time} and the resulting asymmetries are shown in Fig.~\ref{fig:Asym}. 

No result for the \Bz lifetime is quoted, since it is affected by possible biases due to acceptance corrections. These acceptance effects do not influence the measurement of \dmd.

\section{Systematic uncertainties}
\label{sec:systematics}

As explained in Sect.~\ref{sec:acceptance}, systematic effects due to decay time resolution are expected to be small. This is tested using samples of simulated events that are generated with decay time distributions given by the result of the fit to data and convolved with the average measured decay time resolution of $\unit[0.05]{ps}$. The event samples are then fitted with the PDF described in Sect.~\ref{sec:fitmodel}, with the decay time resolution parameter fixed either to zero or to $\sigma_t = \unit[0.10]{ps}$. The maximum observed bias on \dmd of $\unit[0.0002]{ps^{-1}}$ is assigned as systematic uncertainty. Systematic effects due to decay time acceptance are estimated in a similar study, generating samples of simulated events according to the nominal decay time acceptance functions described in Sect.~\ref{sec:acceptance}. These samples are then fitted with the PDF described in Sect.~\ref{sec:fitmodel}, but neglecting the decay time acceptance function in the fit. The average observed shift of $\unit[0.0004]{ps^{-1}}$ ($\unit[0.0001]{ps^{-1}}$) in \BdToDmpip (\BdToJPsiKst) decays is taken as systematic uncertainty. The influence of event-by-event variation of the decay time resolution is found to be negligible.

In order to estimate systematic effects due to the parametrisation of the decay time PDFs for signal and background, an alternative parametrisation is derived with a data-driven method, using \sWeights \cite{external:splot} from a fit to the mass distribution. The \textit{sWeighted} decay time distributions for the signal and background components are then described by Gaussian kernel PDFs, which replace the exponential terms of the decay time PDF. This leads to a description of the data which is independent of a model for the decay time and its acceptance, that can be used to fit for \dmd. The resulting shifts of $\unit[0.0037]{ps^{-1}}$ ($\unit[0.0022]{ps^{-1}}$) in the decay \BdToDmpip (\BdToJPsiKst)  are taken as the systematic uncertainty due to the fit model.

Uncertainties in the geometric description of the detector lead to uncertainties in the measurement of flight distances and the momenta of final state particles. From alignment measurements on the vertex detector, the relative uncertainty on the length scale is known to be smaller than $\unit[0.1]{\%}$. This uncertainty translates directly into a relative systematic uncertainty on \dmd, yielding an absolute uncertainty of $\unit[0.0005]{ps^{-1}}$.

From measurements of biases in the reconstructed \jpsi mass in several run periods, the relative uncertainty on the uncalibrated momentum scale is measured to be smaller than $\unit[0.15]{\%}$. This uncertainty, however, cancels to a large extent in the calculation of the \Bz decay time, as it affects both the reconstructed \Bz momentum and its reconstructed mass, which is dominated by the measured momenta of the final state particles. The remaining systematic uncertainty on the decay time is found to be an order of magnitude smaller than that due to the length scale and is neglected.

A summary of the systematic uncertainties can be found in Table \ref{tab:sys_unc}. The systematic uncertainty on the combined \dmd result is calculated using a weighted average of the combined uncorrelated uncertainties in both channels. The uncertainty on the length scale is fully correlated across the channels and therefore added after the combination. 
\begin{table}[!htb]
\begin{center}
\label{tab:sys_unc}
\caption{Systematic uncertainties on $\Delta m_d$ in $\text{ps}^{-1}$}
\begin{tabular}{lcc}
&$\BdToJPsiKst$&$\BdToDmpip$\\
\hline
Acceptance 		&$0.0001$&$0.0004$\\
Decay time resolution&$0.0002$&$0.0002$\\
Fit model	&$0.0022$&$0.0037$\\
\hline
Total uncorrelated	&$0.0022$&$0.0037$\\
\hline
Length scale 			&$0.0005$&$0.0005$\\
\hline
Total	including correlated &$0.0023$&$0.0037$\\
\end{tabular}
\end{center}
\end{table}
\section{Conclusion}
\label{sec:conclusion}
The \Bd--\Bdb oscillation frequency \dmd has been measured using samples of \BdToDmpip and \BdToJPsiKst events collected in $\unit[1.0]{\invfb}$ of $pp$ collisions at $\sqrt{s} = \unit[7]{TeV}$ and is found to be
\begin{align*}
	\nonumber\dmd(\BdToDmpip) 	&= \unit[0.5178 \pm 0.0061\,(\mathrm{stat.}) \pm 0.0037\,(\mathrm{syst.})]{ps^{-1}}\ \mathrm{and}\\
	\nonumber\dmd(\BdToJPsiKst) &= \unit[0.5096 \pm 0.0114\,(\mathrm{stat.}) \pm 0.0022\,(\mathrm{syst.})]{ps^{-1}}.
\end{align*}
The combined value for \dmd is calculated as the weighted average of the individual results taking correlated systematic uncertainties into account
\begin{align*}
\dmd = \unit[0.5156 \pm 0.0051\,(\mathrm{stat.}) \pm 0.0033\,(\mathrm{syst.})]{ps^{-1}}.
\end{align*}
It is currently the most precise measurement of this parameter. The relative uncertainty on \dmd is  $\unit[1.2]{\%}$, where it is around $\unit[0.6]{\%}$ for \dms \cite{internal:dms}. Thus, the uncertainty on the ratio ${\dmd}/{\dms}$ is dominated by \dmd. As the systematic uncertainties in the \dmd and \dms measurements are small, the error on the ratio can be further improved with more data.

\section*{Acknowledgements}

\noindent We express our gratitude to our colleagues in the CERN
accelerator departments for the excellent performance of the LHC. We
thank the technical and administrative staff at the LHCb
institutes. We acknowledge support from CERN and from the national
agencies: CAPES, CNPq, FAPERJ and FINEP (Brazil); NSFC (China);
CNRS/IN2P3 and Region Auvergne (France); BMBF, DFG, HGF and MPG
(Germany); SFI (Ireland); INFN (Italy); FOM and NWO (The Netherlands);
SCSR (Poland); ANCS/IFA (Romania); MinES, Rosatom, RFBR and NRC
``Kurchatov Institute'' (Russia); MinECo, XuntaGal and GENCAT (Spain);
SNSF and SER (Switzerland); NAS Ukraine (Ukraine); STFC (United
Kingdom); NSF (USA). We also acknowledge the support received from the
ERC under FP7. The Tier1 computing centres are supported by IN2P3
(France), KIT and BMBF (Germany), INFN (Italy), NWO and SURF (The
Netherlands), PIC (Spain), GridPP (United
Kingdom). We are thankful for the computing resources put at our
disposal by Yandex LLC (Russia), as well as to the communities behind
the multiple open source software packages that we depend on.

\addcontentsline{toc}{section}{References}
\bibliographystyle{LHCb}
\bibliography{main}

\end{document}